\documentclass[10pt,letterpaper]{article}
\usepackage{opex3}
\usepackage{color}
\usepackage{amsmath}

\begin{document}

\title{Pulse-compression ghost imaging lidar via coherent detection}

\author{Chenjin Deng, Wenlin Gong$^*$ and Shensheng Han}

\address{Key Laboratory for Quantum Optics and Center for Cold Atom Physics, Shanghai Institute of Optics and Fine Mechanics, Chinese Academy of Sciences, Shanghai 201800, China}

\email{$^*$gongwl@siom.ac.cn} 



\begin{abstract}
Ghost imaging (GI) lidar, as a novel remote sensing technique, has been receiving increasing interest in recent years. By combining pulse-compression technique and coherent detection with GI, we propose a new lidar system called pulse-compression GI lidar. Our analytical results, which are backed up by numerical simulations, demonstrate that pulse-compression GI lidar can obtain the target's spatial intensity distribution, range and moving velocity. Compared with conventional pulsed GI lidar system, pulse-compression GI lidar, without decreasing the range resolution, is easy to obtain high single pulse energy with the use of a long pulse, and the mechanism of coherent detection can eliminate the influence of the stray light, which can dramatically improve the detection sensitivity and detection range.
\end{abstract}



\section{Introduction}
Ghost imaging (GI) is a novel non-scanning imaging method to obtain a target's image with a single-pixel bucket detector \cite{Angelo,Cao,Zhang,Ferri,Gatti,Gong,Shapiro}. Due to the capability of high detection sensitivity, GI has a great application prospect in remote sensing and some kinds of GI imaging lidar system have been proposed \cite{Zhao,Chen,Hardy,Zhu}. Recently, a pulsed three-dimensional (3D) GI lidar was invented and high-resolution 3D image of a natural scene at about 1.0 km range was reported \cite{Gong1}. In this system, the range image was obtained by using simple pulse ranging method, while the azimuth images were reconstructed by computing intensity fluctuation-correlation function between the receiving signal and reference spatial intensity distribution. Because pulsed 3D GI lidar employs direct energy detection, it requires both high peak power and high single pulse energy to obtain sufficient signal-to-noise ratio (SNR). What's more, the range resolution of pulsed 3D GI lidar is determined by the laser's pulse width. In order to obtain high range resolution, it requires a laser with shorter pulse width and a detector with boarder response bandwidth, which usually means that the transmitting system's single pulse energy will be relative low for pulsed 3D GI lidar with a high pulse repetition frequency (PRF). However, for pulsed 3D GI lidar, the detection range mainly depends on single pulse energy, thus high range resolution and long detection range can not be simultaneously achieved.

Coherent detection and pulse-compression technique is valid to solve the conflict described above in chirped amplitude modulated (Chirped-AM) lidar \cite{Stann,Peter}. A Chirped-AM light is emitted, and the return light is received by coherent detection and pulse-compression. Based on this technique, high range resolution and long detection range can be obtained simultaneously \cite{Allen}. Meanwhile, Chirped-AM lidar can achieve the velocity of a moving target \cite{Yu}. Therefore, if we add chirped modulation to pseudo-thermal light source and use coherent detection method to gain the signals reflecting from targets, it is possible to propose a new GI lidar (called pulse-compression GI lidar) with better abilities, which may overcome the difficulties faced with pulsed GI lidar. The paper is organized as follows: in Section II, the system setup and theoretical scheme, including the signal model, light propagation, signal detection, image reconstruction and correction method, is presented; after that, in section III, the numerical results are presented to back up our theoretical result and some discussions on our proposed GI lidar and conventional pulsed 3D GI lidar are given; in Section IV, the conclusion is made.

\section{System setup and analytical results}
\begin{figure}[h]
\centering\includegraphics[width=9cm]{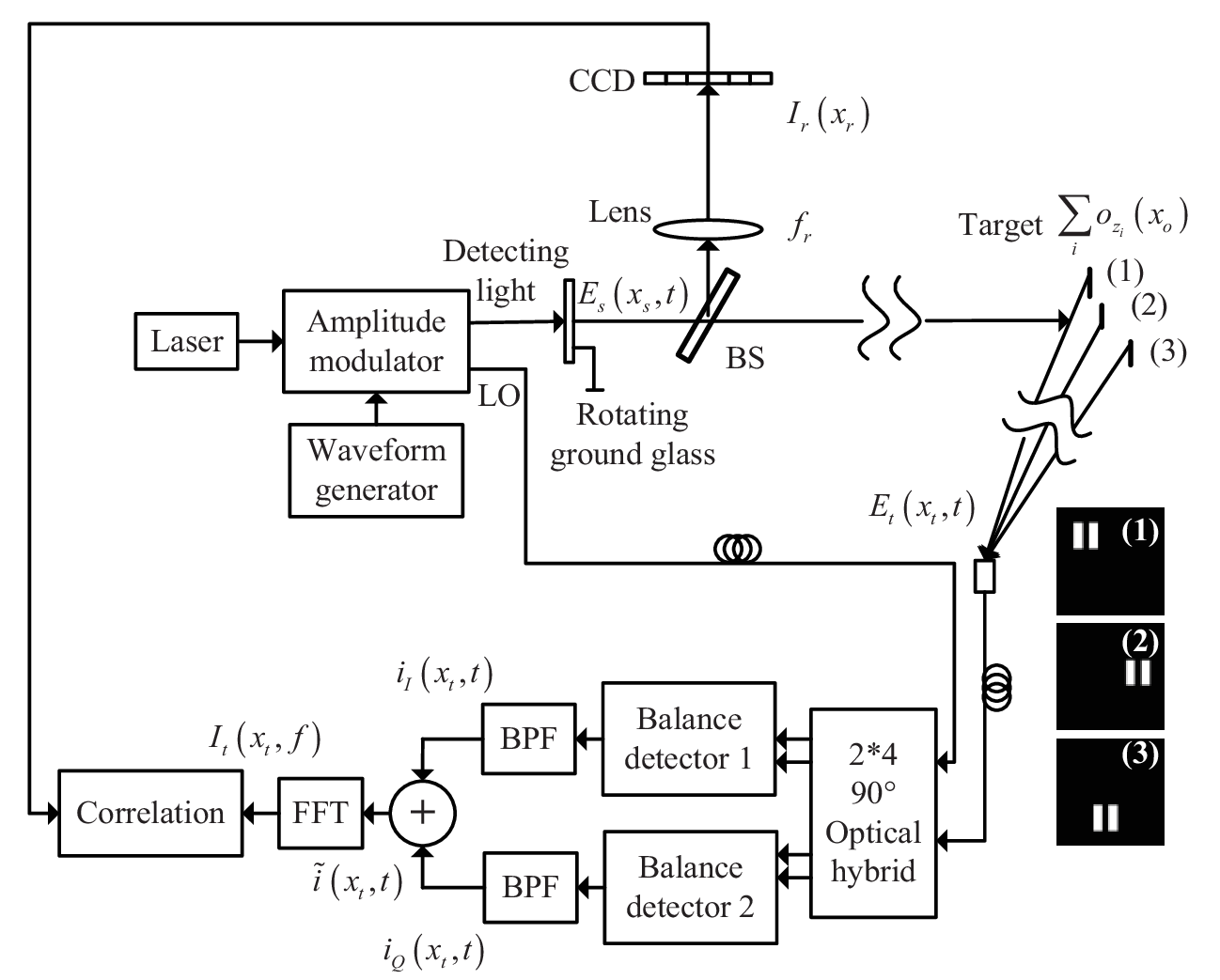}
\caption{The schematic of pulse-compression ghost imaging lidar via coherent detection.}
\end{figure}

Fig. 1 shows the schematic of pulse-compression GI lidar. The laser source with the wavelength $\lambda=1550$ nm is connected to an amplitude modulator and a waveform generator provides the chirped waveform $s\left( t \right)$ for the modulator. The chirped-AM light is split into two parts: detecting light and local oscillator (LO). The spatiotemporal modulated light ${{E}_{s}}\left( {{x}_{s}},t \right)$ is obtained by passing the AM detecting light through a rotating ground glass. Then the light is divided by a beam splitter (BS) into a reference and a test paths. In the reference path, the light is transformed into the far field by an $f_r$-$f_r$ optical system and the spatial intensity distribution ${{I}_{r}}\left( {{x}_{r}} \right)$ is recorded by a charge-coupled device (CCD) camera. In the test path, the light illuminates a 3D target located in the far region of source (namely the rotating ground glass plane). The backscattered light from the target is coupled into a optical fiber and then mixed with LO light in a 2*4 optical hybrid. The four output ports of the optical hybrid are connected to two balanced detectors where de-chirping is performed. After two bandpass filters (BPF), the complex de-chirped current ${{\tilde{i}}}\left({{x}_{t}},t \right)$ is frequency-analyzed by a Fast Fourier transform (FFT) process and the corresponding intensity spectrum ${{I}_{t}}\left( {{x}_{t}},f \right)$ can be obtained.

According to GI theory \cite{Angelo,Cao,Zhang,Ferri,Gatti,Gong}, the intensity fluctuation correlation function between the distribution ${{I}_{r}}\left( {{x}_{r}} \right)$ and the intensity spectrum ${{I}_{t}}\left( {{x}_{t}},f \right)$ can be expressed as
\begin{eqnarray}
\Delta {{G}^{\left( 2,2 \right)}}\left( {{x}_{r}},{{x}_{t}},f \right)=\left\langle {{I}_{r}}\left( {{x}_{r}} \right){{I}_{t}}\left( {{x}_{t}},f \right) \right\rangle -\left\langle {{I}_{r}}\left( {{x}_{r}} \right) \right\rangle \left\langle {{I}_{t}}\left( {{x}_{t}},f \right) \right\rangle,
\end{eqnarray}
where $<\cdot>$ denotes ensemble average of a function.

Since the temporal bandwidth of chirped waveform $s\left( t \right)$ is narrow relative to the optical frequency, the spatiotemporal modulated light field can be treated as a quasi-monochromatic, classical scalar wave \cite{Shapiro2}. The complex envelope of the spatiotemporal light through the rotating ground glass can be described as
\begin{equation}
{{E}_{s,n}}\left( {{x}_{s}},t \right)=\left[ 1+ms\left( t-nT \right) \right]P\left( t-nT \right){{E}_{s,n}}\left( {{x}_{s}} \right),
\end{equation}
where $m$ is the modulation depth, $n$ denotes the $nth$ pulse, and $T$ is the pulse interval; $P\left( t \right)$ is simple pulse waveform and is limited in ${{T}_{0}}$ (${{T}_{0}}<T$); ${{E}_{s,n}}\left( {{x}_{s}} \right)={{A}_{n}}\left( {{x}_{s}} \right)\exp \left[ j{{\phi }_{n}}\left( {{x}_{s}} \right) \right]$ is the spatial amplitude and phase modulation with the following statistical moment
\begin{equation}
~\left\langle {{E}_{s,n}}\left( {{x}_{s}} \right)E_{s,n}^{*}\left( {{{{x}'}}_{s}} \right) \right\rangle ={{I}_{0}}\delta \left( {{x}_{s}}-{{{{x}'}}_{s}} \right),
\end{equation}
where ${{x}_{s}}$ denotes the transverse coordinate at the rotating ground glass plane, ${{I}_{0}}$ is a constant and $\delta \left( x \right)$ is the Dirac's delta function. The chirped waveform $s\left( t \right)$ is
\begin{equation}
s\left( t \right)=\cos \left( 2\pi {{f}_{0}}t+\frac{\pi B{{t}^{2}}}{{{T}_{0}}} \right),
\end{equation}
where ${{f}_{0}}$ is the starting frequency and $B$ is the temporal bandwidth of chirped waveform.

The propagation of light field is described by extended Huygens Fresnel principle \cite{Goodman}. For the optical system depicted in Fig. 1, because the reference CCD camera records all light intensity during the pulse duration, the light intensity distribution ${{I}_{n}}\left( {{x}_{r}} \right)$ is
\begin{equation}
\begin{aligned}
  & \ \ \ {{I}_{n}}\left( {{x}_{r}} \right)\ \propto \int{dt}{{\left| \int{d{{x}_{s}}{{E}_{s,n}}\left( {{x}_{s}},t-\frac{2{{f}_{r}}}{c}-nT \right)\exp \left( \frac{j2\pi {{x}_{r}}{{x}_{s}}}{\lambda {{f}_{r}}} \right)} \right|}^{2}} \\
 & \ \ \ \ \ \ \ \ \ \ \ \ =\int{dt}{{\left[ 1+ms\left( t-\frac{2{{f}_{r}}}{c}-nT \right) \right]}^{2}}{{P}^{2}}\left( t-\frac{2{{f}_{r}}}{c}-nT \right){{\left| \int{d{{x}_{s}}{{E}_{s,n}}\left( {{x}_{s}} \right)\exp \left( \frac{j2\pi {{x}_{r}}{{x}_{s}}}{\lambda {{f}_{r}}} \right)} \right|}^{2}} \\
 & \ \ \ \ \ \ \ \ \ \ \ \ \propto {{\left| \int{d{{x}_{s}}{{E}_{s,n}}\left( {{x}_{s}} \right)\exp \left( \frac{j2\pi {{x}_{r}}{{x}_{s}}}{\lambda {{f}_{r}}} \right)} \right|}^{2}},
\end{aligned}
\end{equation}
where ${{f}_{r}}$ is the focal length of the lens and $x_r$ denotes the transverse coordinate at the CCD camera plane.

In the test path, the light illuminates the target at range ${{z}_{i}}$, and the backscattered light of the target propagates to the receiving aperture plane. The light field at the target plane ${{z}_{i}}$ is
\begin{eqnarray}
{{E}_{o,{{z}_{i}},n}}\left( {{x}_{o}},t \right)=\frac{\exp \left( jk{{z}_{i}} \right)}{j\lambda {{z}_{i}}}\int{d{{x}_{s}}}{{E}_{s,n}}\left( {{x}_{s}},t-\frac{{{z}_{i}}}{c} \right)\ \exp \left[ \frac{j\pi {{\left( {{x}_{o}}-{{x}_{s}} \right)}^{2}}}{\lambda {{z}_{i}}} \right],
\end{eqnarray}
where ${{x}_o}$ is the transverse coordinate at the target plane. And for the target at range ${{z}_{i}}$, its backscattered light field at the receiving aperture plane is
\begin{eqnarray}
{{E}_{t,{{z}_{i}}.n}}\left( {{x}_{t}},t \right)=\frac{\exp \left( jk{{z}_{i}} \right)}{j\lambda {{z}_{i}}}\int{d{{x}_{o}}}{{E}_{o,{{z}_{i}},n}}\left( {{x}_{o}},t-\frac{{{z}_{i}}}{c} \right){{o}_{{{z}_{i}}}}\left( {{x}_{o}} \right)\exp \left[ \frac{j\pi {{\left( {{x}_{t}}-{{x}_{o}} \right)}^{2}}}{\lambda {{z}_{i}}} \right],
\end{eqnarray}
where ${{x}_{t}}$ denotes the transverse coordinate at the receiving aperture and ${{o}_{{{z}_{i}}}}\left( {{x}_{o}} \right)$ is the average reflection coefficient of planar target at the target plane ${{z}_{i}}$.

As depicted in Fig. 1, the target is modeled as a set of quasi planar, spatial extended objects that are located at discrete range ${{z}_{i}}$, and the distance ${{z}_{i}}$ satisfies that Min$\left( {{z}_{i}} \right)>{D_{s}^{2}}/{\lambda}$ (namely in the far field of the source), where ${{D}_{s}}$ is the transverse size of the laser beam on the ground glass plane, and Min$\left( {{z}_{i}} \right)$ is the minimum distance between the target and the source. Moreover, the light illuminating the planar object at certain range cannot reach the object on the plane behind, which means those planar objects have no transverse overlap. Then the total light filed at the receiving aperture plane is given by
\begin{eqnarray}
\begin{aligned}
  & \ \ \ {{E}_{t,n}}\left( {{x}_{t}},t \right)=\sum\limits_{i}{{{E}_{t,{{z}_{i}},n}}\left( {{x}_{t}},t \right)}=\sum\limits_{i}{\frac{\exp \left( j2k{{z}_{i}} \right)}{{{\left( j\lambda {{z}_{i}} \right)}^{2}}}} \\
 & \ \ \ \ \ \ \ \ \ \ \ \ \ \ \ \ \times \int{d{{x}_{s}}}\int{d{{x}_{o}}}{{E}_{s,n}}\left( {{x}_{o}},t-\frac{2{{z}_{i}}}{c} \right)\exp \left[ \frac{j\pi {{\left( {{x}_{o}}-{{x}_{s}} \right)}^{2}}}{\lambda {{z}_{i}}} \right]{{o}_{{{z}_{i}}}}\left( {{x}_{o}} \right)\exp \left[ \frac{j\pi {{\left( {{x}_{t}}-{{x}_{o}} \right)}^{2}}}{\lambda {{z}_{i}}} \right].
\end{aligned}
\end{eqnarray}

If the target moves along the optical axis, the range can be described as ${{z}_{i}}={{z}_{{{i}_{0}}}}+{{v}_{i}}t$, where ${{z}_{{{i}_{0}}}}$ is the range at $t=0$, and ${{v}_{i}}$ is the radial velocity. Due to the high PRF, the sampling time is so short that the target can be supposed to stay in a range resolution cell for simplicity. Then Eq. (8) becomes
\begin{eqnarray}
{{E}_{t,n}}\left( {{x}_{t}},t \right)=\sum\limits_{i}{\left[ 1+ms\left( t-\frac{2{{z}_{{{i}_{0}}}}}{c}-nT \right) \right]}P\left( t-\frac{2{{z}_{{{i}_{0}}}}}{c}-nT \right)\exp \left( j2\pi {{f}_{{{d}_{i}}}}t \right){{E}_{t,n,i}}\left( {{x}_{t}} \right),
\end{eqnarray}
where ${{f}_{{{d}_{i}}}}={2{{v}_{i}}}/{\lambda }\;$ is the Doppler frequency and ${{E}_{t,n,i}}\left( {{x}_{t}} \right)$ denotes as
\begin{eqnarray}
\begin{aligned}
  & {{E}_{t,n,i}}\left( {{x}_{t}} \right)={{A}_{t,n,i}}\exp \left[ j{{\phi }_{t,n,i}}\left( {{x}_{t}} \right) \right]\equiv \sum\limits_{i}{\frac{\exp \left( j2k{{z}_{{{i}_{0}}}} \right)}{{{\left( j\lambda {{z}_{{{i}_{0}}}} \right)}^{2}}}} \\
 & \ \ \ \ \ \ \ \ \ \ \ \ \times \int{d{{x}_{s}}}\int{d{{x}_{o}}}{{E}_{s,n}}\left( {{x}_{o}},t-\frac{2{{z}_{{{i}_{0}}}}}{c} \right)\exp \left[ \frac{j\pi {{\left( {{x}_{o}}-{{x}_{s}} \right)}^{2}}}{\lambda {{z}_{{{i}_{0}}}}} \right]{{o}_{{{z}_{i}}}}\left( {{x}_{o}} \right)\exp \left[ \frac{j\pi {{\left( {{x}_{t}}-{{x}_{o}} \right)}^{2}}}{\lambda {{z}_{{{i}_{0}}}}} \right],
\end{aligned}
\end{eqnarray}

The LO light filed is assumed to be uniform, namely
\begin{eqnarray}
{{E}_{LO,n}}\left( {{x}_{t}},t \right)=\left[ 1+ms\left( t-nT \right) \right]P\left( t-nT \right){{A}_{LO}}\exp \left[ j{{\phi }_{LO,n}} \right],
\end{eqnarray}
where ${{A}_{LO}}$ and ${{\phi }_{LO,n}}$ is the amplitude and the known phase of the $nth$ LO pulse, respectively. In coherent detection system, the signal light must be spatially coherent at the receiving aperture to obtain maximum mixing efficiency. Suppose the transverse scale of the target is L, and then the transverse coherent length of light field on receiving aperture is $\sim {\lambda {{z}_{i}}}/{L}\;$, which means our receiver size should not exceed this constraint \cite{Protopopov}. The 2*4 optical hybrid mixes the signal light with four quadrature states associated with the LO light field, and then delivers the four light signals to two balanced detectors. The interference items of in-phase (I) and quadrature (Q) channels can be written as \cite{Kazovsky}
\begin{eqnarray}
\begin{aligned}
  & I:\ \ \ \ 2\left[ {{E}_{t,n}}\left( {{x}_{t}},t \right)E_{LO,n}^{*}\left( {{x}_{t}},t \right)+E_{t,n}^{*}\left( {{x}_{t}},t \right){{E}_{LO,n}}\left( {{x}_{t}},t \right) \right] \\
 & Q:\ \ \ 2\left[ {{E}_{t,n}}\left( {{x}_{t}},t \right)E_{LO,n}^{*}\left( {{x}_{t}},t \right)\exp \left( -\frac{j\pi }{2} \right)+E_{t,n}^{*}\left( {{x}_{t}},t \right){{E}_{LO,n}}\left( {{x}_{t}},t \right)\exp \left( \frac{j\pi }{2} \right) \right].
\end{aligned}
\end{eqnarray}

Optical mixing and de-chirping process happens simultaneously in balanced detectors. By using two proper BPFs, we can achieve the oscillating component of the photocurrent that varies harmonically with Doppler frequency and range frequency. The output currents are
\begin{eqnarray}
\begin{aligned}
  & \ {{i}_{I,n}}\left( {{x}_{t}},t \right)=\sum\limits_{i}{\left\{ 1+{{m}^{2}}\cos \left[ {4\pi \beta {{z}_{{{i}_{0}}}}t}/{c}\;+{{\phi }_{i0,n}} \right] \right\}P\left( t-nT \right)} \\
 & \ \ \ \ \ \ \ \ \ \ \ \ \ \ \ \ \ \ \ \ \ \times {{A}_{LO}}{{A}_{t,n,i}}\left( {{x}_{t}} \right)\cos \left[ 2\pi {{f}_{d}}t+{{\phi }_{t,n,i}}\left( {{x}_{t}} \right)-{{\phi }_{LO,n}} \right] \\
 & {{i}_{Q,n}}\left( {{x}_{t}},t \right)=\sum\limits_{i}{\left\{ 1+{{m}^{2}}\cos \left[ {4\pi \beta {{z}_{{{i}_{0}}}}t}/{c}\;+{{\phi }_{i0,n}} \right] \right\}P\left( t-nT \right)} \\
 & \ \ \ \ \ \ \ \ \ \ \ \ \ \ \ \ \ \ \ \ \ \times {{A}_{LO}}{{A}_{t,n,i}}\left( {{x}_{t}} \right)\sin \left[ 2\pi {{f}_{d}}t+{{\phi }_{t,n,i}}\left( {{x}_{t}} \right)-{{\phi }_{LO,n}} \right],
\end{aligned}
\end{eqnarray}
where ${{\phi }_{i0,n}}={4\pi {{f}_{o}}{{z}_{{{i}_{0}}}}}/{c}\;-{4\pi \beta {{z}_{{{i}_{0}}}}nT}/{c}\;-\pi \beta {{\left( {2{{z}_{{{i}_{0}}}}}/{c}\; \right)}^{2}}$, and the complex output current is
\begin{eqnarray}
{{\tilde{i}}_{n}}\left( {{x}_{t}},t \right)={{i}_{I,n}}\left( {{x}_{t}},t \right)+j*{{i}_{Q,n}}\left( {{x}_{t}},t \right).
\end{eqnarray}
After FFT process, the intensity spectrum of ${{\tilde{i}}_{n}}\left( {{x}_{t}},t \right)$ is
\begin{eqnarray}
{{I}_{t,n}}\left( {{x}_{t}},f \right)=\sum\limits_{i}{\left\{ {{\operatorname{sinc}}^{2}}\left[ T\left( f-{{f}_{{{d}_{i}}}} \right) \right]\text{+}{{m}^{2}}{{\operatorname{sinc}}^{2}}\left[ T\left( f-{{f}_{{{d}_{i}}}}-\frac{2{{z}_{{{i}_{0}}}}\beta }{c} \right) \right] \right\}{{\left| {{A}_{LO}} \right|}^{2}}{{\left| {{A}_{t,n,i}}\left( {{x}_{t}} \right) \right|}^{2}}}.
\end{eqnarray}
where $\sin c(x)=\frac{\sin(\pi x)}{\pi x}$.

If the targets shown in Fig. 1 have surfaces that are sufficiently rough (on the scale of an optical wavelength), then $\left\langle {o_{zi}(x)o^{\ast}_{zi}(x')} \right\rangle=O(x)\delta (x-x' )$ \cite{Hardy}. Substituting Eqs. (5), (10) and (15) into Eq. (1), and suppose the field fluctuations obey a complex circular Gaussian random process with zero mean \cite{Goodman1}, after some calculation, we can get
\begin{eqnarray}
\begin{aligned}
  & \Delta {{G}^{\left( 2,2 \right)}}\left( {{x}_{r}},{{x}_{t}},f \right)\propto {{\left| {{A}_{LO}} \right|}^{2}}\sum\limits_{i}{\left\{ {{\operatorname{sinc}}^{2}}\left[ T\left( f-{{f}_{{{d}_{i}}}} \right) \right]\text{+}{{m}^{2}}{{\operatorname{sinc}}^{2}}\left[ T\left( f-{{f}_{{{d}_{i}}}}-\frac{2{{z}_{{{i}_{0}}}}\beta }{c} \right) \right] \right\}} \\
 & \ \ \ \ \ \ \ \ \ \ \ \ \ \ \ \ \ \ \ \ \ \ \ \times \int{d{{x}_{o}}}{{O}_{{{z}_{i}}}}\left( {{x}_{o}} \right){{\operatorname{sinc}}^{2}}\left[ \frac{{{D}_{s}}\left( {{x}_{r}}-{{{f}_{r}}{{x}_{o}}}/{{{z}_{{{i}_{0}}}}}\; \right)}{\lambda {{f}_{r}}} \right].
\end{aligned}
\end{eqnarray}
Eq. (16) suggests that the angular resolution is ${\lambda }/{{{D}_{s}}}$ and the range resolution is ${c}/{2B}$. Meanwhile, the target's information can be extracted with the use of a single point-like detector when the measurement process reaches ensemble average. However, in practice, in order to obtain a proper information output rate, the measurement number is usually small and coherent detection efficiency experiences significant degradation due to the fluctuation of the backscattered light field, thus the visibility of GI with a single coherent detector is very poor \cite{Wang}. Following Ref. \cite{Chan}, we can employ a random sparse coherent detection array to improve the detection SNR by summing the restored intensity spectrums, which is equivalent to increasing the measurement number and the visibility of GI will be enhanced.

Moreover, if the target contains moving planar object, Eq. (16) implies that the range frequency is blurred with Doppler frequency, i.e. ${{f}_{{{b}_{i}}}}={{f}_{{{d}_{i}}}}+{2{{z}_{{{i}_{0}}}}\beta }/{c}\;$. Based on Doppler frequency and the recovered tomographic images, we try to obtain the correct range information. The most important step is to determine whether the reconstructed image is from a static scatter or a moving scatter. According to Eq. (16), the corresponding tomographic images in different frequency $f$ can be independently reconstructed. Because the target doesn't overlap in transverse dimension, if there is no overlap for all tomographic images at ${{x}_{r}}={{x}_{{{r}_{o}}}}$, then the scatter is static, whereas if there is overlap for two tomographic images at ${{x}_{r}}={{x}_{{{r}_{o}}}}$, then the corresponding scatter is moving. The corrected range and velocity can be expressed as
\begin{equation}
\begin{aligned}
  & z=c\left( {{f}_{b}}-{{f}_{d}} \right)/2\beta  \\
 & v={\lambda {{f}_{d}}}/{2}\;, \\
\end{aligned}
\end{equation}
Therefore, the target's spatial intensity distribution, range and moving velocity can be achieved by pulse-compression GI lidar.

\section{Simulation results and Discussion}
\begin{figure}[h]
\centering\includegraphics[width=7cm]{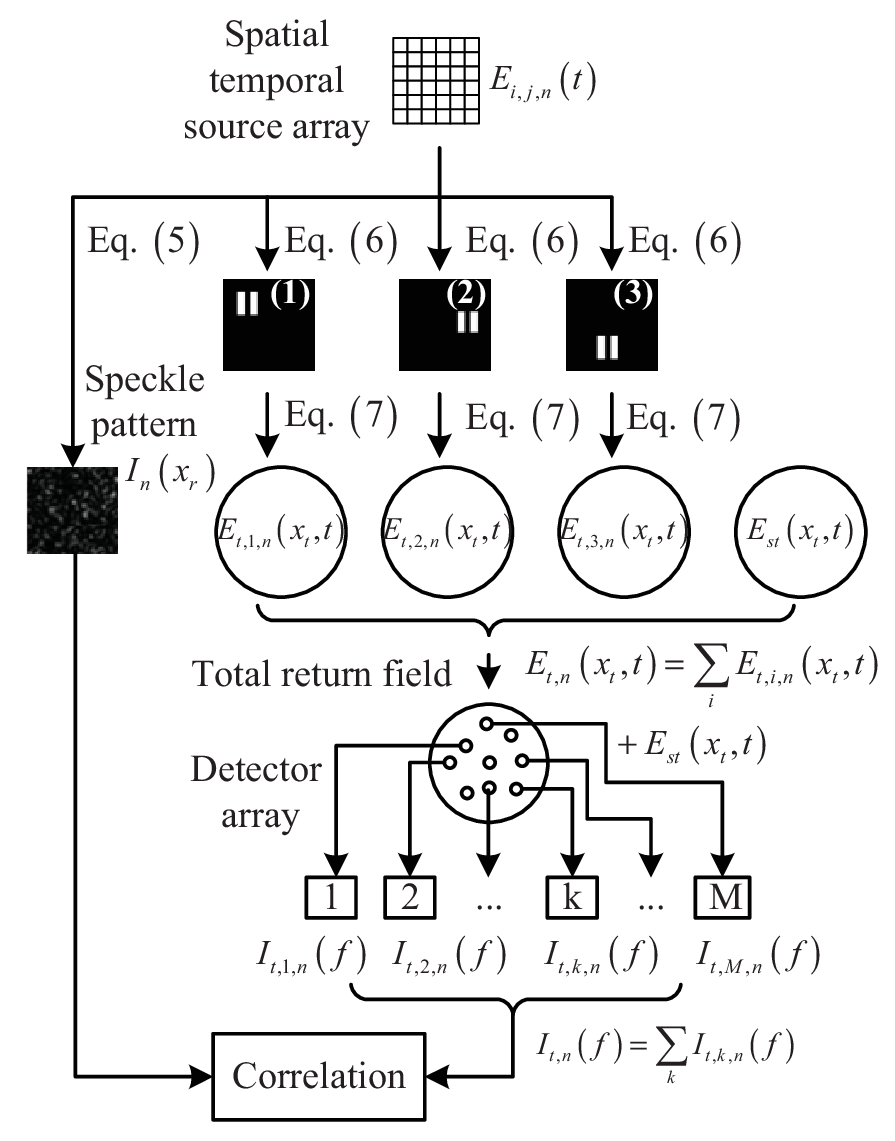}
\caption{Schematic of simulation and process.}
\end{figure}

In order to verify the analytical results of our proposed pulse-compression GI lidar above, we give some numerical simulations and Fig. 2 shows the simulation process. The spatial temporal source is discretized as a two-dimensional (2D) lattices, namely ${{E}_{i,j}}\left( t \right)=\left[ 1+ms\left( t \right) \right]{{A}_{i,j}}\exp \left[ j{{\phi }_{i,j}} \right]$. Following Ref. \cite{Zhang1}, the amplitude and phase are statistically independent of each other. And all element sources are independent identically distributed. The amplitude ${{A}_{i,j}}$ obeys Rayleigh distribution and the phase ${{\phi }_{i,j}}$ is uniformly distributed on $\left( 0,\ 2\pi  \right)$.

The reference intensity distribution ${{I}_{r,n}}\left( {{x}_{r}} \right)$ is obtained by computing Eq. (5). In the test path, the target's reflection function is discretized as a set of 2D lattices, corresponding to different planar objects. Since the planar objects do not occlude each other, the propagation of each planar object can be computed independently. The spatiotemporal light fields ${{E}_{o,{{z}_{i}},n}}\left( {{x}_{o}},t \right)$ and ${{E}_{t,{{z}_{i}},n}}\left( {{x}_{t}},t \right)$ are achieved by computing the numerical result of Eq. (6) and Eq. (7), respectively. To demonstrate our Lidar's performance in scenarios with stray light, a random stray light field ${{E}_{st}}\left( {{x}_{t}},t \right)$ is generated at receiving aperture plane. Following Eq. (9), the total field ${{E}_{t,n}}\left( {{x}_{t}},t \right)$ at the receiving aperture is the coherent superposition of all planar objects' return filed and stray light.

The detection process is simulated by computing Eq. (12) and (13). By using a digital BPF, we can get the baseband current ${{\tilde{i}}_{n}}\left( {{x}_{t}},t \right)$, corresponding to Eq. (14). To simulate a random sparse detection array, we randomly pick up some positions on the receiving aperture plane, as shown in Fig. 2. After FFT process, the intensity spectrum ${{I}_{t,k,n}}\left( {{x}_{t}},f \right)$ can be obtained, where $k$ denotes the $kth$ detector at ${{x}_{t}}$. At last, the image of pulse-compression GI lidar is reconstructed by computing the following correlation function
\begin{eqnarray}
\Delta {{G}^{\left( 2,2 \right)}}\left( {{x}_{r}},f \right)=\frac{1}{N}\sum\limits_{n}{{{I}_{r,n}}\left( {{x}_{r}} \right)\left[ \sum\limits_{k}{{{I}_{t,k,n}}\left( {{x}_{t}},f \right)} \right]}-\left[ \frac{1}{N}\sum\limits_{n}{{{I}_{r,n}}\left( {{x}_{r}} \right)} \right]\left\{ \frac{1}{N}\sum\limits_{n}{\left[ \sum\limits_{k}{{{I}_{t,k,n}}\left( {{x}_{t}},f \right)} \right]} \right\},
\end{eqnarray}
where $N$ denotes the total measurement number.

In the numerical simulations, the specific parameters are set as follows: ${{D}_{s}}$=2mm, $T$=1ms, $m$=1, and $B$=1GHz. As shown in Fig. 1, the three planar objects are the identical double slit (slit width $a$=0.5m, slit height $h$=1.5m, and center-to-center separation $d$=0.87m) at different ranges (object 1 at 199.9m, object 2 at 200m and object 3 at 200.3m) with different transverse positions. For the scenario with moving components, object 1 and 2 has a radial velocity 0.1 m/s and 0.2 m/s respectively, while object 3 is static.

\begin{figure}[h]
\centering\includegraphics[width=7cm]{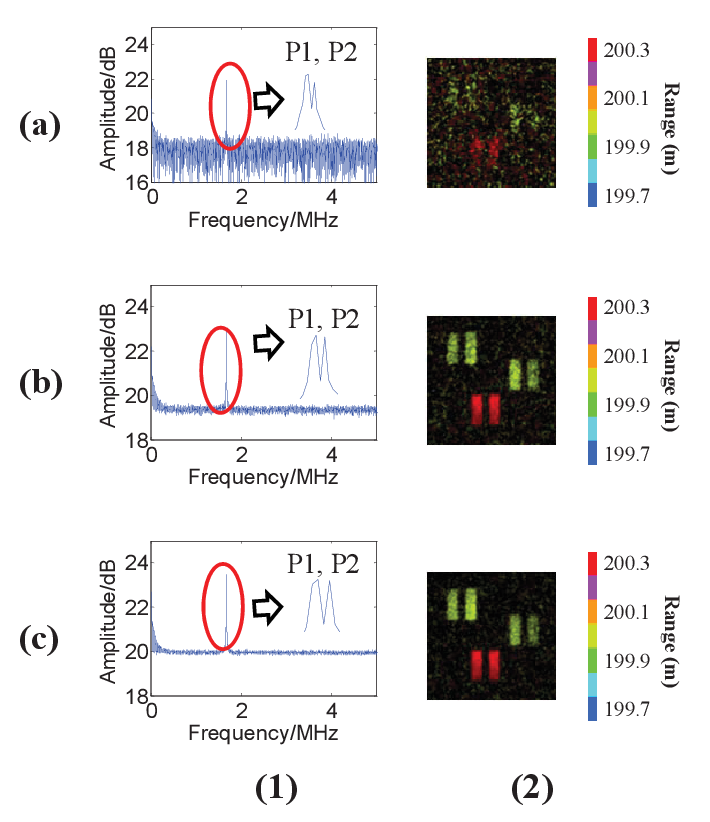}
\caption{Simulation results of pulse-compression GI lidar for a static scenario (consist of three planar objects). (a), (b) and (c) are the intensity spectrum and image reconstruction results by using 1, 25, and 100 coherent receivers, respectively (averaged 20000 measurements); Column (1) and (2) present the intensity spectrum, GI reconstruction results for peak frequency component P1 and P2, respectively.}
\end{figure}

\begin{figure}[h]
\centering\includegraphics[width=7cm]{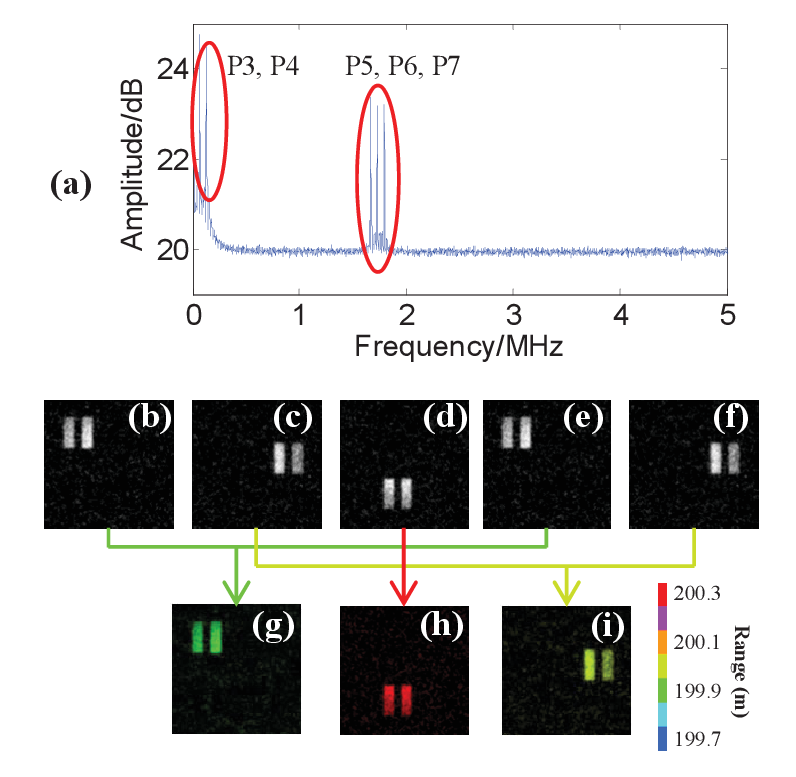}
\caption{Simulation results of pulse-compression GI lidar for a scenario with moving objects. The setup of the target is the same as simulation 1 except for that planar object 1 and 2 has a radial velocity 0.1 m/s and 0.2 m/s, respectively. (a) is the intensity spectrum of the random sparse detection array with 100 detectors, the labeled circles on the left and the right are the Doppler frequency region and coupling frequency region, respectively; (b)-(f) are GI reconstruction results for the peak frequency components P3-P7 (averaged 20000 measurements); (g)-(i) are the reconstructed image of the object 1, object 3 and object 2, respectively.}
\end{figure}

\begin{figure}[h]
\centering\includegraphics[width=7cm]{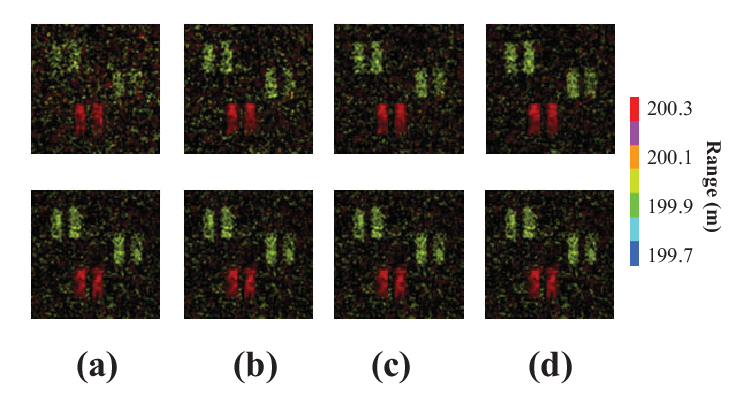}
\caption{Simulation results of conventional pulsed 3D GI lidar and pulse-compression GI lidar in different level of stay light (with 25 random sparse detectors and averaged 5000 measurements). The upper line is the recovered results of conventional pulsed 3D GI lidar and the bottom line is the restored results obtained by pulse-compression GI lidar. (a) SNR=1 dB; (b) SNR=3 dB; (c) SNR=5 dB; (d) SNR=10 dB.}
\end{figure}

Fig. 3 presents the results of pulse-compression GI lidar for a static scenario. Column (1) is the intensity spectrum of the random sparse detection array with 1, 10 and 100 detectors. By computing Eq. (19), the corresponding reconstruction images for the labeled peak frequency components P1 and P2 are illustrated in Column (2). As illustrated in Fig. 3(a), the intensity spectrum's SNR for a single coherent detector is low and so does the GI reconstruction result. However, when some random sparse receivers are used to collect the backscattered light from the target, both the intensity spectrum's SNR and the reconstruction quality of GI dramatically increase with the number of receivers. In addition, as predicted by the theory, the range resolution of pulse-compression GI lidar is ${c}/{2B}$=0.15m, thus the spectrum of the object 1 and object 2 cannot be resolved and as shown in Column (2), both the object 1 and object 2 appear in the same tomographic image.

When image a scenario with moving objects, Fig. 4(a) presents the intensity spectrum of the random sparse detection array with 100 detectors. The peaks P3 and P4 are corresponding to the Doppler frequency and the peaks P5-P7 are the blurred range frequency components. The corresponding reconstructed images for the peak frequency components P3-P7 are shown in Figs. 4(b)-(f). According to the image correction process described above, we can identify that Fig. 4(b) and Fig. 4(e) correspond to the same planar object (namely the image of object 1), Fig. 4(d) is a static object (namely the image of object 3), while Fig. 4(c) and Fig. 4(f) are the image of object 2. In addition, by computing Eq. (18), the correct ranges and velocities are ${{z}_{1}}=199.88m$, ${{v}_{1}}=0.0996{m}/{s}$, ${{z}_{2}}=199.97m$, ${{v}_{2}}=0.199{m}/{s}$ and ${{z}_{1}}=200.28m$. Therefore, the range resolution can be also enhanced if the Doppler information of the object is used.

To illustrate the performance of pulse-compression GI lidar in scenarios with stray light, we carry out a comparison between conventional pulsed GI lidar and pulse-compression GI lidar. Using the same simulation parameters of Fig. 3 and 25 random sparse detectors, Fig. 5 gives the reconstruction results of conventional pulse GI lidar and pulse-compression GI lidar when the detection SNR for a single detector is 1dB, 3dB, 5dB, and 10 dB, respectively. It is clearly seen that the reconstruction quality of pulse GI lidar increases with the detection SNR. However, pulse-compression GI lidar hardly depends on the detection SNR because the LO light does't interfere with the stray light, which means that pulse-compression GI lidar can eliminate the influence of the stray light.

\section{Conclusion}
In summary, we demonstrate by theoretical analysis and numerical simulation that coherent detection and pulse compression can be applied in GI lidar to image a 3D scenario with moving components. The emitting laser is spatiotemporally modulated, and the received pulse is de-chirped in optical domain using coherent detection. The proposed pulse-compression GI lidar uses low peak power pulse and low sampling rate to obtain high range resolution. Compared with conventional pulsed 3D GI lidar, pulse-compression GI lidar can effectively eliminate the influence of the stray light to the imaging quality, which is very useful to weak remote sensing.

\section{Acknowledgments}
The work was supported by the Hi-Tech Research and Development Program of China under Grant Project No. 2013AA122901, Natural Science Foundation of China under Grant Project No. 61571427, and Youth Innovation Promotion Association CAS No. 2013162.
\end{document}